\documentclass{optica-article}

\journal{opticajournal} 

\articletype{Research Article}

\usepackage{graphicx}
\usepackage{bm}

\newcommand\mat\mathbf
\newcommand\ii{{\rm i}}

\newcommand\dd{\mathop{}\!{\rm d}}

\newcommand\nm{\mathrm{\;nm}}
\newcommand\um{\mathrm{\;\mu m}}
\newcommand\uminv{\mathrm{\;\mu m}^{-1}}

\begin{document}

\title{Optical computation of the divergence of a vector field using a metal-dielectric multilayer}

\author{%
Leonid~L.~Doskolovich,\authormark{1,2,*} 
Artem~I.~Kashapov,\authormark{1,2}
Evgeni~A.~Bezus,\authormark{1,2} and 
Dmitry~A.~Bykov\authormark{1,2}%
}

\address{%
\authormark{1}Image Processing Systems Institute, NRC ``Kurchatov Institute'', 
151 Molodogvardeyskaya st., Samara 443001, Russia\\
\authormark{2}Samara National Research University, 
34 Moskovskoye shosse, Samara 443086, Russia%
}

\email{\authormark{*}leonid@ipsiras.ru}

\begin{abstract*}
We theoretically describe the optical computation of the divergence of a two-dimensional vector field, which is composed by the transverse electric field components of an incident light beam. The divergence is computed in reflection at oblique incidence of light on a layered structure. We show that in the particular case of a linearly polarized incident beam, the layered structure implementing the divergence operator also allows one to compute the gradient and perform the isotropic differentiation. As an example of a layered structure computing the divergence, we propose a metal-dielectric multilayer consisting of two pairs of metal and dielectric layers on a metal substrate. The presented numerical simulation results of the designed multilayer confirm that the divergence operator is computed with high accuracy. We also demonstrate the possibility of using the designed structure for optical directional differentiation and computation of the gradient and Laplace operators.
\end{abstract*}

\section{Introduction}

Nowadays, the development of photonic structures for analog optical computing attracts considerable research interest. Such structures performing computations ``at the speed of light'' are considered as promising building blocks for next-generation computing systems, in which photonic components will serve as an alternative to their electronic counterparts or supplement them in task-specific applications~\cite{1,2,3}. Much attention in the field of analog optical computing is paid to the development of compact photonic structures for differentiating optical signals (with respect to time or spatial coordinates, as well as in the spatiotemporal domain)~\cite{3,4,5,6,7,8,9,10,11,12,13,14,15,16,17,18,19,20,21,22}. In addition to the differentiation, of great interest is the optical implementation of various differential operators~\cite{2,23,24,25,26,27,28}, for example, the Laplace operator~\cite{2,23,24,25,26}, which can be used for all-optical edge detection.

One of the basic differential operators in vector calculus is the divergence operator. Optical computation of this operator is, in particular, useful for the analysis of the structure of cylindrical vector beams and the determination of polarization singularities~\cite{27}. At the same time, to the best of our knowledge, there exist only two theoretical works dedicated to the optical implementation of this operator~\cite{27,28}.
Let us first discuss, what is understood under the optical computation of the divergence using a photonic structure. Let us consider an incident light beam, in which the polarization at each point of the plane perpendicular to the beam propagation direction is linear. With such a beam, one can associate a two-dimensional vector field composed of the transverse components of the electric field. To implement the divergence operator, a photonic structure has to generate (either reflected or transmitted) light beam, in which one of the transverse electric field components is proportional to the divergence of the specified two-dimensional vector field.
In~\cite{27}, a rather complex structure consisting of a phase plate and a resonant plasmonic diffraction grating with a graphene layer was proposed for the optical computation of the divergence operator.
The divergence operator was computed in reflection due to the simultaneous utilization of the resonant excitation of the plasmonic mode of the grating and of the optical analogue of the spin Hall effect.
In~\cite{28}, a dielectric tetrahedron with metal layers deposited on two of its faces was proposed for computing the divergence.
In this case, the divergence computation was performed due to plasmonic resonances occurring at successive reflections of the incident beam from the faces of the tetrahedron.
Unfortunately, in~\cite{28}, no numerical simulation results were presented to demonstrate the quality of computing this operator.

In this work, we theoretically describe the all-optical computation of the divergence operator using a layered structure.
Similarly to~\cite{27}, the computation of the divergence operator is based on the joint utilization of the optical resonance effect, which ensures the appearance of a reflection zero, and the optical analogue of the spin Hall effect. 
We show that in the particular case of a linearly polarized incident beam with a single nonzero transverse electric field component, 
the layered structure implementing the divergence operator also allows one to compute the gradient. 
In this case, the structure also implements the so-called isotropic differentiation, in which the intensity of the reflected beam is proportional to the squared modulus of the gradient. 
As an example of a layered structure implementing these operations, we propose a metal-dielectric multilayer consisting of two pairs of metal and dielectric layers on a metal substrate.
The presented numerical simulation results of the designed multilayer show the computation of the divergence operator and the gradient with high quality.
We also demonstrate that the same structure can be used for the optical computation of the Laplacian.
In the opinion of the present authors, the four-layer structure proposed for implementing the divergence operator is significantly simpler compared to the structure from~\cite{27}, which consists of a resonant plasmonic grating with a graphene layer and a phase plate. 
The proposed structure is also significantly more compact compared to the bulky tetrahedron-based structure from~\cite{28}.

The present work is organized as follows.
In Section~\ref{sec:1}, a theoretical description of the diffraction of a three-dimensional light beam on a layered structure is presented and a vectorial transfer function describing the transformation of the electric field components of the incident beam occurring upon reflection from the structure is obtained. 
In Section~\ref{sec:2}, we present the conditions required for the optical computation of the divergence operator in one of the transverse components of the reflected beam. 
Section~\ref{sec:3} shows that for a linearly polarized incident beam with a single nonzero transverse electric field component, 
a layered structure satisfying these conditions also enables computing the gradient. 
Sections~\ref{sec:4} and~\ref{sec:5} are dedicated to the design and numerical investigation of a metal-dielectric multilayer implementing the divergence operator. Section~\ref{sec:6} concludes the paper.

\section{\label{sec:1}Diffraction of a three-dimensional optical beam on a layered structure}

\subsection{\label{sec:1:1}Representation of the incident beam}

\begin{figure}
\centering\includegraphics{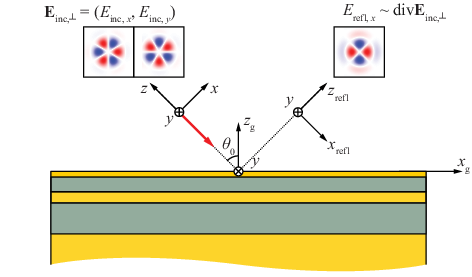}
\caption{\label{fig:1} Diffraction of a three-dimensional optical beam on a layered structure.
The red arrow shows the direction of the incident beam.
}
\end{figure}

Let us consider a monochromatic light beam obliquely incident on a layered structure at an angle $\theta_0$. 
For the further derivations, it is convenient to work
in a coordinate system $(x, y, z)$ associated with the incident beam and chosen so that the beam propagates in the negative direction of the $z$ axis (Fig.~\ref{fig:1}).
This coordinate system is rotated by the angle $\theta_0$ around the $y$ axis with respect to the ``global'' coordinate system $(x_{\rm g}, y, z_{\rm g})$ associated with the layered structure.
The coordinates of the vectors in the two considered coordinate systems are related by the rotation matrix $\mat{R}_y$ as
\begin{equation}\label{eq:1} 
\begin{bmatrix}
 x \\ 
 y \\ 
 z \\
\end{bmatrix}
= 
\mat{R}_y
\begin{bmatrix}
 x_{\rm g} \\ 
 y \\ 
 z_{\rm g} \\
\end{bmatrix}
=
\begin{bmatrix}
    \cos\theta_0 & 0 & \sin\theta_0 \\ 
   0    			 & 1 & 0  				\\
   -\sin\theta_0 & 0 & \cos\theta_0 \\
 \end{bmatrix}
\begin{bmatrix}
 x_{\rm g} \\ 
 y \\ 
 z_{\rm g} \\
\end{bmatrix}.
\end{equation}

In order to describe the diffraction of the incident beam on the layered structure, it is convenient to represent it in a plane wave basis.
For this, we will use plane waves, which have transverse magnetic (TM) and transverse electric (TE) polarizations in the global coordinate system $(x_{\rm g},y,z_{\rm g})$.
These TM- (TE-) polarized plane waves have the magnetic (electric) field vector perpendicular to the plane containing the wave vector and the normal to the surface of the layered structure (the $z_\mathrm{g}$ axis).
In the coordinate system associated with the incident beam, the electric field components $E_x$, $E_y$, and $E_z$ of the TM- and TE-polarized plane waves can be represented as~\cite{13, 26}
\begin{equation}\label{eq:2}
\mat{E}_{\rm TM,TE}(\mat{r}, \mat{k}) = \mat{A}_{\rm TM,TE} (\mat{k}_{\bot}) \exp \left\{ \ii \mat{k} \cdot \mat{r} \right\} 
 = \mat{A}_{\rm TM,TE} (\mat{k}_{\bot}) \exp \left\{ \ii \mat{k}_{\bot} \cdot \mat{r}_{\bot} +\ii k_z z \right\},
\end{equation}
where $\mat{r} = (x, y, z)^{\rm T}$ and $\mat{k} = (k_x, k_y, k_z)^{\rm T}$ is the wave vector of the plane wave.
The transverse components of these vectors are denoted by
$\mat{r}_{\bot} = (x, y)^{\rm T}$ and $\mat{k}_{\bot} = (k_x, k_y)^{\rm T}$.
The vectors $\mat{A}_{\rm TM,TE}(\mat{k}_{\bot})$ in Eq.~\eqref{eq:2} contain the amplitudes of the $E_x$, $E_y$, and $E_z$ components of the TM- and TE-polarized plane waves, respectively.
From the Maxwell's equations and Eq.~\eqref{eq:1}, it follows that they have the form~\cite{13}
\begin{equation}\label{eq:3}
\begin{aligned}
\mat{A}_{\rm TM}(\mat{k}_{\bot}) 
&=
 \frac{1}{\gamma}
\begin{bmatrix}
    - k_x k_z \cos \theta_0 + (k_y^2 + k_z^2) \sin\theta_0  \\ 
    - k_y (k_z \cos \theta_0 + k_x \sin\theta_0)  \\ 
     (k_x^2 + k_y^2) \cos \theta_0 - k_x k_z \sin\theta_0  \\ 
\end{bmatrix}
,
\\
\mat{A}_{\rm TE}(\mat{k}_{\bot})
&=
\frac{1}{\gamma}
\begin{bmatrix}
  -k_0  k_y \cos\theta_0 \\ 
   k_0 (k_x \cos\theta_0 - k_z \sin\theta_0) \\ 
   k_0  k_y \sin\theta_0  \\ 
 \end{bmatrix},
\end{aligned}
\end{equation}
where
\begin{equation}\label{eq:gamma}
\gamma = 
\sqrt{(k_x \cos\theta_0 - k_z \sin\theta_0)^2 + k_y^2}.
\end{equation}
Let us note that the components of the wave vector $\mat{k}$ in Eqs.~\eqref{eq:2} and~\eqref{eq:3} satisfy the dispersion relation of the plane wave $k_x^2 + k_y^2 + k_z^2 = k_0^2 \varepsilon_{\rm sup}$, where $k_0 = 2\pi/\lambda$ is the wave number, $\lambda$ is the free-space wavelength, and $\varepsilon_{\rm sup}$ is the dielectric permittivity of the medium. 
Let us also note that we consider the vectors $\mat{A}_{\rm TM,TE}(\mat{k}_{\bot})$ in Eqs.~\eqref{eq:2} and~\eqref{eq:3} 
as functions of only the transverse wave vector components $\mat{k}_{\bot} = (k_x, k_y)^\mathrm{T}$, since the remaining component is expressed through them as $k_z = \pm \sqrt{k_0^2\varepsilon_{\rm sup} - \mat{k}_{\bot}^2}$.
Here, the plus and minus signs are used to describe waves propagating in positive and negative directions of the $z$~axis, respectively. 

Having written the equations of the TE- and TM-polarized plane waves, we can represent the incident beam as their superposition:
\begin{equation}\label{eq:5}
  \mat{E}_{\rm inc}(\mat{r}) 
	=
	\iint \left[G_{\rm TM}(\mat{k}_{\bot}) \mat{A}_{\rm TM} (\mat{k}_{\bot}) 
	+ G_{\rm TE}(\mat{k}_{\bot})\mat{A}_{\rm TE}(\mat{k}_{\bot}) \right]
	\exp \left\{ 
		 \ii \mat{k}_{\bot} \cdot \mat{r}_{\bot} 
		-\ii z \sqrt{k_0^2 \varepsilon_{\rm sup} - \mat{k}_{\bot}^2} 
	\right\}
	\dd{\mat{k}_{\bot}},
\end{equation}
where $G_{\rm TM}(\mat{k}_{\bot})$ and $G_{\rm TE}(\mat{k}_{\bot})$ are the spectra of the incident beam representing the amplitudes of the TM- and TE-polarized waves, respectively.
Let us note that since the incident beam propagates in the negative direction of the $z$~axis, the wave vector component $k_z$ in Eq.~\eqref{eq:5} was taken with a minus sign, i.\,e., $k_z = -\sqrt{k_0^2 \varepsilon_{\rm sup} - \mat{k}_{\bot}^2}$.
The same expression for $k_z$ has to be used in Eqs.~\eqref{eq:3} and~\eqref{eq:gamma} defining the amplitudes $\mat{A}_{\rm TM}(\mat{k}_{\bot})$ and $\mat{A}_{\rm TE}(\mat{k}_{\bot})$.

In the general case, by choosing the spectra $G_{\rm TM}(\mat{k}_{\bot})$ and $G_{\rm TE}(\mat{k}_{\bot})$, one can ensure that two electric field components of the incident beam have a certain required form.
Let us assume that in the plane $z = 0$, the transverse electric field components $E_{{\rm inc},x}(\mat{r}_{\bot},0)$ and $E_{{\rm inc},y}(\mat{r}_{\bot},0)$ are known and are defined through their spectra $G_x(\mat{k}_{\bot})$ and $G_y(\mat{k}_{\bot})$ as
\begin{equation}\label{eq:6}
E_{{\rm inc},x}(\mat{r}_{\bot}, 0) 
= \iint G_x(\mat{k}_{\bot}) \exp \{\ii \mat{k}_{\bot} \cdot\mat{r}_{\bot}\} \dd{\mat{k}_{\bot}},
\end{equation}
\begin{equation}\label{eq:7}
E_{{\rm inc},y}(\mat{r}_{\bot}, 0) 
= \iint G_y(\mat{k}_{\bot}) \exp \{\ii\mat{k}_{\bot} \cdot\mat{r}_{\bot}\} \dd{\mat{k}_{\bot}}.
\end{equation}
Using Eqs.~\eqref{eq:5} and~\eqref{eq:3}, one can easily show that the spectra $G_{\rm TM}(\mat{k}_{\bot})$ and $G_{\rm TE}(\mat{k}_{\bot})$, which describe the incident field distribution, can be expressed through $G_x(\mat{k}_{\bot})$ and $G_y(\mat{k}_{\bot})$ as
\begin{equation}\label{eq:8}
\begin{aligned}
  G_{\rm TM}(\mat{k}_{\bot}) 
	&=g_{{\rm TM},x}(\mat{k}_{\bot}) G_x(\mat{k}_{\bot}) 
	 + g_{{\rm TM},y}(\mat{k}_{\bot}) G_y(\mat{k}_{\bot}),
	\\
  G_{\rm TE}(\mat{k}_{\bot}) 
	&=g_{{\rm TE},x}(\mat{k}_{\bot}) G_x(\mat{k}_{\bot}) 
	 + g_{{\rm TE},y}(\mat{k}_{\bot}) G_y(\mat{k}_{\bot}), 
\end{aligned}
\end{equation}
where
\begin{equation}\label{eq:9:1}
  g_{{\rm TM}, x}(\mat{k}_{\bot}) 
	= \frac{-k_x \cos\theta_0 + k_z \sin\theta_0}
	        { k_z \gamma},
\end{equation}
\begin{equation}\label{eq:9:2}
	g_{{\rm TM}, y}(\mat{k}_{\bot}) 
	= \frac{-k_y \cos\theta_0}
	        { k_z \gamma},
\end{equation}
\begin{equation}\label{eq:9:3}
  g_{{\rm TE}, x}(\mat{k}_{\bot}) 
	= -\frac{k_y(k_x \sin \theta_0 + k_z \cos\theta_0)}
					 {k_0 k_z \gamma},
\end{equation}
\begin{equation}\label{eq:9:4}
	g_{{\rm TE}, y}(\mat{k}_{\bot}) 
	= \frac{k_x k_z \cos \theta_0 + (k_x^2 - k_0^2 \varepsilon_{\rm sup})\sin \theta_0 }
	        {k_0 k_z \gamma}.
\end{equation}

\subsection{Transformation of the beam upon reflection from a layered structure}

When the incident beam is reflected from a layered structure, the amplitudes of the TM- and TE-polarized plane waves constituting the beam are multiplied by the reflection coefficients $R_{\rm TM}(\theta)$ and $R_{\rm TE}(\theta)$, respectively. 
These coefficients depend on the acute angle of incidence $\theta$, which is the angle between the wave vector of the plane wave and the normal to the surface of the layered structure.
Using Eq.~\eqref{eq:1}, one can show that this angle depends on $\mat{k}_{\bot} =(k_x,k_y)^{\rm T}$ as
\begin{equation}
\label{eq:theta}
\theta(\mat{k}_{\bot}) = \arcsin \sqrt{\frac{(k_x \cos\theta_0 + \sqrt{k_0^2 \varepsilon_{\rm sup} - \mat{k}_{\bot}^2} \sin\theta_0)^2 + k_y^2}{k_0^2 \varepsilon_{\rm sup}}}.
\end{equation}

In the coordinate system of the reflected beam $(x_{\rm refl}, y, z_{\rm refl})$ (Fig.~\ref{fig:1}), the transverse components of the wave vectors of the incident waves $\mat{k}_{\bot}$ do not change upon reflection, and the component $k_z$ changes sign, since the reflected beam propagates in the positive direction of the $z_{\rm refl}$ axis.
Thus, taking into account Eq.~\eqref{eq:8}, we can write the reflected beam in the considered coordinate system as
\begin{equation}\label{eq:10}
\mat{E}_{\rm refl}(\mat{r}_{\rm refl}) 
= \iint \left[ G_x(\mat{k}_{\bot}) \bm{\Theta}_x(\mat{k}_{\bot}) 
             + G_y(\mat{k}_{\bot}) \bm{\Theta}_y(\mat{k}_{\bot}) \right] 
\exp \left\{ 
	\ii\mat{k}_{\bot}\cdot\mat{r}_{\bot, {\rm refl}} + 
	\ii z_{\rm refl} \sqrt{k_0^2 \varepsilon_{\rm sup} - \mat{k}_{\bot}^2}  
\right\}
\dd{\mat{k}_{\bot}},
\end{equation}
where $\mat{r}_{{\rm refl}} = (x_{\rm refl}, y, z_{\rm refl})^{\rm T}$, $\mat{r}_{\bot, {\rm refl}} = (x_{\rm refl}, y)^{\rm T}$, and
\begin{equation}\label{eq:11}
\begin{aligned}
  \bm{\Theta}_x  (\mat{k}_{\bot}) 
	&= g_{{\rm TM},x}(\mat{k}_{\bot}) \mat{A}_{\rm TM}(\mat{k}_{\bot}) R_{\rm TM}(\theta(\mat{k}_{\bot})) 
	+ g_{{\rm TE},x}(\mat{k}_{\bot}) \mat{A}_{\rm TE}(\mat{k}_{\bot}) R_{\rm TE}(\theta(\mat{k}_{\bot})),
	\\
  \bm{\Theta}_y(\mat{k}_{\bot}) 
	&= g_{{\rm TM},y}(\mat{k}_{\bot}) \mat{A}_{\rm TM}(\mat{k}_{\bot}) R_{\rm TM}(\theta(\mat{k}_{\bot})) 
	+ g_{{\rm TE},y}(\mat{k}_{\bot}) \mat{A}_{\rm TE}(\mat{k}_{\bot}) R_{\rm TE}(\theta(\mat{k}_{\bot})).
	\end{aligned}
\end{equation}
Here, the functions $g_{{\rm TM}, x}(\mat{k}_{\bot}),\, g_{{\rm TM}, y}(\mat{k}_{\bot})$ and $g_{{\rm TE}, x}(\mat{k}_{\bot}),\, g_{{\rm TE}, y}(\mat{k}_{\bot})$ are defined by Eqs.~\eqref{eq:9:1}--\eqref{eq:9:4}, and the vectors $\mat{A}_{\rm TM}(\mat{k}_{\bot})$ and $\mat{A}_{\rm TE}(\mat{k}_{\bot})$ have the form of Eq.~\eqref{eq:3} with $k_z = +\sqrt {k_0^2 \varepsilon_{\rm sup} - \mat{k}_{\bot}^2}$ and $\theta_0$ replaced with $-\theta_0$.

From Eqs.~\eqref{eq:6},~\eqref{eq:10}, and~\eqref{eq:11}, it follows that the electric field components of the reflected beam at $z_{\rm refl} = 0$ correspond to the sum of two transformations of the transverse electric field components of the incident beam $E_{{\rm inc}, x}(\mat{r}_{\bot},0)$ and $E_{{\rm inc}, y}(\mat{r}_{\bot},0)$ performed by linear systems with vectorial transfer functions (TFs) $\bm{\Theta}_x(\mat{k}_{\bot})$ and $\bm{\Theta}_y(\mat{k}_{\bot})$ defined by Eq.~\eqref{eq:11}.

\section{\label{sec:2}Theoretical conditions for computing the divergence operator}

In this section, we will obtain the conditions necessary for the optical computation of the divergence of a two-dimensional vector field consisting of the transverse electric field components of the incident beam $E_{{\rm inc},x}(\mat{r}_{\bot},0)$ and ${E_{{\rm inc},y}}({\mat{r}_{ \bot}},0)$. We will assume that the spectra of the incident beam $G_x(\mat{k}_{ \bot})$ and $G_y(\mat{k}_{ \bot})$ are ``concentrated'' in the vicinity of the point $\mat{k}_{ \bot} = (0,0)$,
which corresponds to $\theta = \theta_0$.
In this case, the TFs of Eq.~\eqref{eq:11} in this vicinity can be approximated by their Taylor series expansions up to linear terms.
Using Eq.~\eqref{eq:11} with Eqs.~\eqref{eq:3},~\eqref{eq:9:1}--\eqref{eq:9:4}, and~\eqref{eq:theta}, we obtain these expansions as 
\begin{equation}\label{eq:12}
\bm{\Theta}_x(\mat{k}_{\bot}) 
\approx
\begin{bmatrix}
   R_{\rm TM}(\theta_0) \\ 
   0  \\ 
   0 
\end{bmatrix}
+
\begin{bmatrix}
   c_x  \\ 
   0  \\ 
   -\frac{R_{\rm TM}(\theta_0)}{k_0\sqrt{\varepsilon_{\rm sup}}}
 \end{bmatrix} k_x
+
\begin{bmatrix}
   0  \\ 
   c_y  \\ 
   0 
\end{bmatrix}
k_y,
\end{equation}
\begin{equation}\label{eq:13} 
\bm{\Theta}_y(\mat{k}_{\bot}) 
\approx 
\begin{bmatrix}
   0  \\ 
   R_{\rm TE}(\theta_0) \\ 
   0 
\end{bmatrix}
 +
\begin{bmatrix}
   0  \\ 
   {\tilde c}_x  \\ 
   0 
\end{bmatrix}
k_x
+
\begin{bmatrix}
   c_y \\ 
   0   \\ 
   -\frac{R_{\rm TE}(\theta_0)}{k_0\sqrt{\varepsilon_{\rm sup}} }
\end{bmatrix}
 k_y,
\end{equation}
where
\begin{equation}\label{eq:14} 
c_x = \frac{R_{\rm TM}'(\theta_0)}{k_0\sqrt{\varepsilon_{\rm sup}}} 
, \;\;\;
\tilde c_x = \frac{R_{\rm TE}'(\theta_0)}{k_0\sqrt{\varepsilon_{\rm sup}}} 
, \;\;\;
c_y = 
\frac{ R_{\rm TM}(\theta_0) - R_{\rm TE}(\theta_0)}
		 {k_0\sqrt{\varepsilon_{\rm sup}} }
\cot\theta_0.
\end{equation}
From Eqs.~\eqref{eq:6} and~\eqref{eq:7}, one can easily see that the terms linear with respect to $k_x$ and $k_y$ in the TFs of Eqs.~\eqref{eq:12} and~\eqref{eq:13} describe the differentiation of the transverse electric field components with respect to the spatial variables $x$ and $y$. Thus, in the linear approximation of Eqs.~\eqref{eq:12} and~\eqref{eq:13}, a layered structure implements the following transformations in the electric field components of the reflected beam:
\begin{equation}\label{eq:15}
E_{{\rm refl}, x}(\mat{r}_{\bot}, 0) 
= R_{\rm TM}(\theta_0) E_{{\rm inc}, x}(\mat{r}_{\bot},0) 
- \ii c_x \frac{\partial E_{{\rm inc},x}(\mat{r}_{\bot},0)}
							 {\partial x}
- \ii c_y \frac{\partial E_{{\rm inc},y}(\mat{r}_{\bot},0)}
							 {\partial y},
\end{equation}
\begin{equation}\label{eq:16}
E_{{\rm refl}, y}(\mat{r}_{\bot}, 0) 
= R_{\rm TE}(\theta_0) E_{{\rm inc},y}(\mat{r}_{\bot},0)
- \ii c_y \frac{\partial E_{{\rm inc},x}(\mat{r}_{\bot},0)}
							 {\partial y}
- \ii{\tilde c}_x\frac{\partial E_{{\rm inc},y}(\mat{r}_{\bot}, 0)}
											{\partial x},
\end{equation}
\begin{equation}\label{eq:17}
E_{{\rm refl}, z}(\mat{r}_{\bot}, 0) 
=  \frac{\ii}{k_0\sqrt{\varepsilon_{\rm sup}}}
	  \left( R_{\rm TM}(\theta_0)
\frac{\partial E_{{\rm inc},x}(\mat{r}_{\bot},0)}
		 {\partial x} 
+ R_{\rm TE}(\theta_0)
\frac{\partial E_{{\rm inc},y}(\mat{r}_{\bot},0)}
		 {\partial y} \right).
\end{equation}

Let us consider the particular transformation implemented when the reflection coefficient of the layered structure $R_{\rm TM}(\theta_0)$ vanishes and the coefficients $c_x$ and $c_y$ defined by Eq.~\eqref{eq:14} coincide.
These two conditions can be written as 
\begin{equation}\label{eq:18}
R_{\rm TM}(\theta_0) = 0, \;\;\;
R_{\rm TM}'(\theta_0) = -R_{\rm TE}(\theta_0) \cot\theta_0.
\end{equation}
From Eq.~\eqref{eq:15}, it is evident that in this case, the transformation performed in the $x_{\rm refl}$ electric field component of the reflected beam describes (up to a factor) the computation of the divergence operator applied to a two-dimensional vector field 
$\mat{E}_{{\rm inc}, \bot} = \left( E_{{\rm inc},x}(\mat{r}_{\bot}, 0), E_{{\rm inc},y}(\mat{r}_{\bot},0) \right)$:
\begin{equation}\label{eq:19}
E_{{\rm refl},x}( \mat{r}_{\bot}, 0) 
= -\ii c_x \left( \frac{\partial E_{{\rm inc}, x} (\mat{r}_{\bot}, 0)}
											 {\partial x} + 
									\frac{\partial E_{{\rm inc}, y} (\mat{r}_{\bot}, 0)}
											 {\partial y} \right) 
= -\ii c_x \operatorname{div} \mat{E}_{{\rm inc}, \bot}.
\end{equation}
	
Similarly, if the reflection coefficient of the layered structure $R_{\rm TE}(\theta_0)$ vanishes and the condition ${\tilde c_x} = c_y$ is fulfilled, then the transformation of Eq.~\eqref{eq:16} carried out in the $y$ electric field component of the reflected beam is proportional to the divergence operator of a two-dimensional vector field $\left( E_{{\rm inc},y}(\mat{r}_{ \bot},0), E_{{\rm inc},x}(\mat{r}_{\bot},0) \right)$:
\begin{equation}\label{eq:20}
E_{{\rm refl},y}(\mat{r}_{\bot},0) 
= -\ii{\tilde c_x}\left( 
\frac{\partial E_{{\rm inc}, y}(\mat{r}_{\bot},0)}
     {\partial x}
+ \frac{\partial E_{{\rm inc},x}(\mat{r}_{\bot},0)}
       {\partial y} \right).
\end{equation}
Let us note that the second terms in Eqs.~\eqref{eq:19} and~\eqref{eq:20}, which contain the derivatives of the cross-polarized components $E_{{\rm inc},y}(\mat{r}_{\bot},0)$ and $E_{{\rm inc}, x}(\mat{r}_{\bot}, 0)$, can be considered as a manifestation of the optical analogue of the spin Hall effect~\cite{29,30}.
At the same time, the zero reflection conditions $R_{\rm TM}(\theta_0) = 0$ or $R_{\rm TE}(\theta_0) = 0$ are caused, as a rule, by resonant effects associated with the excitation of eigenmodes of the layered structure being used~\cite{10,11,12,13}.
Thus, the computation of the divergence operator can be performed by simultaneous utilization of the optical analogue of the spin Hall effect and a resonant effect providing a reflection zero.

\section{\label{sec:3}Computation of the gradient for a linearly polarized incident beam}

It is worth noting that layered structures enabling the computation of the divergence operator make it possible to perform other important operations including the computation of the gradient and the so-called ``isotropic'' differentiation. Indeed, let us consider a layered structure satisfying the two conditions of Eq.~\eqref{eq:18}. As it was shown above, such a structure provides, in the linear approximation of Eqs.~\eqref{eq:12} and~\eqref{eq:13}, the computation of the divergence operator in the ${x_{\rm refl}}$-component of the electric field of the reflected beam [see Eq.~\eqref{eq:19}].

Let a linearly polarized beam with the electric field components $E_{{\rm inc},x}(\mat{r}_{\bot}, 0)$ defined by Eq.~\eqref{eq:6} and $E_{{\rm inc}, y} (\mat{r}_{\bot}, 0) \equiv 0$ impinge on a layered structure satisfying the conditions of Eq.~\eqref{eq:18}. In this case, $G_y(\mat{k}_{\bot}) \equiv 0$ and, according to Eqs.~\eqref{eq:10} and~\eqref{eq:11}, the electric field components of the reflected beam $\mat{E}_{\rm refl}(\mat{r}_{\bot, {\rm refl}}, 0)$ will correspond to the transformation of the $x$ electric field component of the incident beam $E_{{\rm inc}, x}(\mat{r}_{\bot}, 0)$ by a linear system with the TF ${\bm{\Theta}_x}(\mat{k}_{\bot})$ having the form of Eq.~\eqref{eq:12}. From Eqs.~\eqref{eq:12} and~\eqref{eq:18}, it is easy to obtain that for an incident beam that is linearly polarized along the $x$ axis, the transverse electric field components of the reflected beam $\mat{E}_{ \bot, {\rm refl}}(\mat{r}_{ \bot}, 0) = \left( E_{{\rm refl}, x}(\mat{r}_{ \bot}, 0), E_{{\rm refl},y}(\mat{r}_{\bot},0) \right)$ will, up to a factor, be equal to the gradient of the function $E_{{\rm inc},x}(\mat{r}_{\bot}, 0)$:
\begin{equation}\label{eq:21}
\mat{E}_{\bot, {\rm refl}}({\mat{r}_{ \bot}},0) 
= - \ii c_x\operatorname{grad} {E_{{\rm inc}, x}}(\mat{r}_{ \bot}, 0) 
= - \ii c_x\left( 
\frac{\partial E_{{\rm inc}, x}(\mat{r}_{\bot}, 0)}
     {\partial x},
\frac{\partial E_{{\rm inc},x}({\mat{r}_{\bot}},0)}
     {\partial y} \right).
\end{equation}
Let us note that in the linear approximation of Eq.~\eqref{eq:12}, the $z_{\rm refl}$-component $E_{{\rm refl}, z}(\mat{r}_{\bot},0)$ will be zero. Therefore, the intensity of the reflected beam $I_{\rm refl} = \left| \mat{E}_{\rm refl} \right|^2 = \left| E_{{\rm refl},x} \right|^2 + \left| E_{{\rm refl},y} \right|^2 + \left| E_{{\rm refl},z} \right|^2$ will be proportional to the squared magnitude of the gradient of the incident electric field component $E_{{\rm inc}, x}$:
\begin{equation}\label{eq:22}
I_{\rm refl}(\mat{r}_{\bot}) 
= |c_x|^2 \left( \left| 
\frac{\partial E_{{\rm inc},x}(\mat{r}_{ \bot},0)}
		 {\partial x} 
								 \right|^2 
+ \left| \frac{\partial E_{{\rm inc},x}(\mat{r}_{ \bot}, 0)}
						  {\partial y} 
 \right|^2 \right) 
= |c_x|^2 \left| \operatorname{grad}E_{{\rm inc},x} (\mat{r}_{\bot}, 0) \right|^2.
\end{equation}
The transformation described by Eqs.~\eqref{eq:21} and~\eqref{eq:22} can be referred to as the isotropic differentiation operation, since it enables enhancing the contours in input images (field distributions) isotropically for all contour orientations.

\section{\label{sec:4}Metal-dielectric layered structure for computing the divergence and gradient operators}

Let us consider the calculation of a layered structure for computing the divergence operator in the $x_{\rm refl}$ electric field component of the reflected beam. In this case, as it was shown in Section~\ref{sec:2}, the two conditions of Eq.~\eqref{eq:18} have to be fulfilled. 
According to the results of the previous section, in addition to divergence, such a layered structure will compute the gradient of the function $E_{{\rm inc}, x}(\mat{r}_{\bot},0)$ in the case of an incident beam, which is linearly polarized along the $x$ axis (and thus has $E_{{\rm inc}, y}(\mat{r}_{\bot},0) \equiv 0$).

Let us note that in the general case, the reflection coefficients $R_{\rm TM}(\theta_0)$ and $R_{\rm TE}(\theta_0)$ of a layered structure and, consequently, their derivatives are complex numbers.
Therefore, the condition $R_{\rm TM}'(\theta_0) = -R_{\rm TE}(\theta_0) \cot\theta_0$ from Eq.~\eqref{eq:18} in fact consists of two equations describing the equality of the real and imaginary parts of complex values in its left- and right-hand sides.
Thus, the layered structure being utilized must not only provide the reflection zero $R_{\rm TM}(\theta_0) = 0$, but also possess at least two free parameters to fulfill the second condition from Eq.~\eqref{eq:18}.

\begin{figure}
\centering\includegraphics{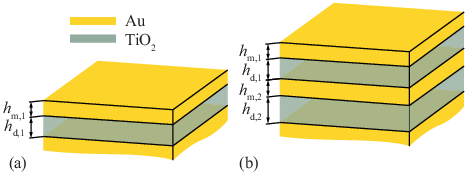}
\caption{\label{fig:2} Geometry of the two-layer metal-dielectric-metal structure~(a)
and of the four-layer structure used for the computation of the divergence operator~(b).}
\end{figure}

We believe that one of the simplest layered structures, in which it is possible to obtain a reflection zero, is the metal-dielectric-metal structure shown in Fig.~\ref{fig:2}(a), which consists of an upper metal layer (with thickness $h_{{\rm m},1}$ and dielectric permittivity $\varepsilon_{{\rm m},1}$) and a dielectric layer (thickness $h_{{\rm d},1}$ and dielectric permittivity $\varepsilon_{\rm d}$) on a metal substrate (an optically thick metal layer with dielectric permittivity $\varepsilon_{\rm sub}$).
The reflection zero in this structure appears due to a Fabry--P{\'e}rot resonance under the so-called critical coupling condition. It is important to note that a reflection zero can be obtained for any fixed parameters of the incident wave (wavelength $\lambda$, angle of incidence $\theta_0$, TM- or TE-polarization) by choosing the thicknesses $h_{{\rm m},1}$ and $h_{{\rm d},1}$ of the metal and dielectric layers~\cite{10,11}. The thickness of the metal layer $h_{{\rm m},1}$ is found by solving the equation
\begin{equation}\label{eq:24}
\left| \frac{r_1}{r_1 r_2 - t^2} \right| = |r_3|,
\end{equation}
where $r_1$ and $r_2$ are the complex reflection coefficients of the metal layer for plane waves incident from above (from the superstrate) and from below (from the dielectric layer), $t$ is the transmittance coefficient of the metal layer, and $r_3$ is the reflection coefficient of the metal substrate. Let us note that the metal layer thickness $h_{{\rm m},1}$ enters this equation implicitly through the coefficients $r_1 = r_1(h_{{\rm m},1})$, $r_2 = r_2(h_{{\rm m},1})$, and $t = t(h_{{\rm m},1})$, which are defined by the well-known formulas describing the reflection and transmission of a plane wave through a homogeneous thin film~\cite{31}. When finding a solution to Eq.~\eqref{eq:24}, the wavelength~$\lambda$, angle of incidence~$\theta_0$, and polarization of the incident wave (TM or TE) as well as the dielectric permittivities of the materials $\varepsilon_{{\rm m},1}$, $\varepsilon_{\rm d}$, $\varepsilon_{\rm sup}$, and $\varepsilon_{\rm sub}$, on which the mentioned coefficients also depend, are considered as fixed parameters. Equation~\eqref{eq:24} is solved numerically and, since only its left-hand side depends on the metal layer thickness $h_{{\rm m},1}$, the solution in the simplest case can be performed by a brute-force search on a certain one-dimensional grid. After the metal layer thickness is found from Eq.~\eqref{eq:24}, the dielectric layer thickness can be calculated analytically as~\cite{10, 11}
\begin{equation}\label{eq:25}
h_{{\rm d},1} = \frac{1}{2 k_0 \sqrt{\varepsilon_{\rm d}-\varepsilon_{\rm sup}\sin^2\theta_0}}\left( 2\pi n + \arg \frac{r_1}{r_3 \left( r_1 r_2 - t^2 \right)} \right),\;\; n \in {\mathbb N}.
\end{equation}

The simple metal-dielectric-metal structure considered above enables obtaining a reflection zero necessary for the computation of the divergence operator, but does not have two free parameters required to fulfill the second necessary condition $c_x = c_y$ described by Eq.~\eqref{eq:18}. Due to this, in order to compute the divergence, we propose to use a four-layer structure containing an additional pair of metal and dielectric layers [Fig.~\ref{fig:2}(b)]. The thicknesses of these layers $h_{{\rm m},2}$ and $h_{{\rm d},2}$ can be considered as the two required free parameters. At fixed thicknesses of the lower layers $h_{{\rm m},2}$ and $h_{{\rm d},2}$, the thicknesses of the upper layers $h_{{\rm m},1}$ and $h_{{\rm d},1}$ providing a reflection zero can still be calculated from Eqs.~\eqref{eq:24} and~\eqref{eq:25}, in which as the reflection coefficient $r_3$, the reflection coefficient of the structure consisting of two lower layers on the metal substrate has to be used.

As a particular example, we chose four-layer ``${\rm Au}-{\rm TiO}_2-{\rm Au}-{\rm TiO}_2$'' multilayers consisting of two pairs of gold and titanium dioxide layers on a gold substrate (optically thick layer) [Fig.~\ref{fig:2}(b)].
In order to find, whether it is possible to fulfill the condition of Eq.~\eqref{eq:18}, the multilayers with different lower layer thicknesses $h_{{\rm m},2}$ and $h_{{\rm d},2}$ possessing a reflection zero $R_{\rm TM}(\theta_0) = 0$ at the free-space wavelength $\lambda = 633\nm$ and angle of incidence $\theta_0 = 25^\circ$ were calculated using Eqs.~\eqref{eq:24} and~\eqref{eq:25}, and then the condition of Eq.~\eqref{eq:18} was checked.
For calculating the reflection coefficients and their derivatives appearing in Eq.~\eqref{eq:18}, the numerically stable enhanced transmittance matrix approach was utilized~\cite{32}.
As the dielectric permittivities of the materials, reference data were used~\cite{33}.
As a result, a metal-dielectric multilayer was found, for which 
$c_x = 0.260 \times \exp\{-2.995\ii\}$, 
$c_y = 0.263 \times \exp\{-2.998\ii\}$, and, therefore, the required condition $c_x = c_y$ is approximately satisfied with a good accuracy. The layer thicknesses of the calculated structure are the following:
\begin{equation}\label{eq:26}
h_{{\rm m}, 1} = 5.2\nm,\;\;
h_{{\rm d}, 1} = 39.6\nm,\;\;
h_{{\rm m}, 2} = 52.5\nm,\;\;
h_{{\rm d}, 2} = 70.7\nm.
\end{equation}

\begin{figure}
\centering\includegraphics{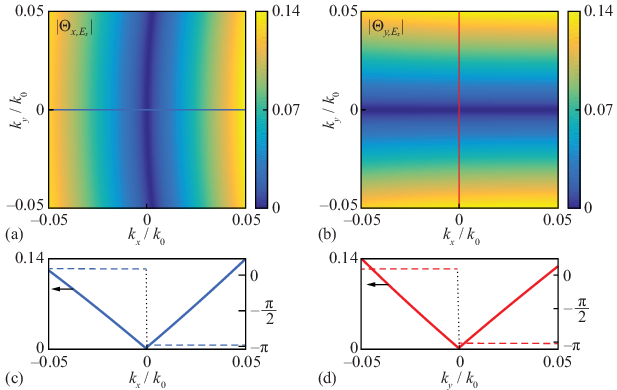}
\caption{\label{fig:3} Absolute values of the calculated TFs $\Theta_{x, E_x}(\mat{k}_{\bot})$~(a) and $\Theta_{y, E_x}(\mat{k}_{\bot})$~(b) describing the formation of the $x_{\rm refl}$ electric field component of the beam reflected from the multilayer of Eq.~\eqref{eq:26}. (c),~(d)~Cross-sections of~(a) and~(b) along the corresponding solid lines.}
\end{figure}

Figure~\ref{fig:3} shows the calculated absolute values of the TFs $\Theta_{x, E_x}(\mat{k}_{\bot})$ and $\Theta_{y, E_x}(\mat{k}_{\bot})$
 of the designed metal-dielectric multilayer of Eq.~\eqref{eq:26}. 
These TFs are the first components of the vectorial TFs of Eq.~\eqref{eq:11} and describe the formation of the $x_{\rm refl}$ electric field component of the reflected beam. Let us note that the calculation of the reflection coefficients in these TFs was also carried out using the numerically stable enhanced transmittance matrix approach~\cite{32}. 
From Fig.~\ref{fig:3}, it is evident that in the considered spatial frequency range $|k_x|/k_0 < 0.05$, $|k_y|/k_0 < 0.05$, the rigorously calculated TFs of Eq.~\eqref{eq:11} are well described by their linear approximations 
$c_x k_x$ and $c_y k_y$ 
 (the root-mean-square deviation of the calculated TFs from their linear approximations does not exceed~2.3\%).

\section{\label{sec:5}Numerical simulation results}

In this section, we will consider several applications of the designed metal-dielectric multilayer [Eq.~\eqref{eq:26}] implementing the divergence operator. 
In Subsection~\ref{sec:5:1}, we use this structure to compute a directional derivative; 
in Subsection~\ref{sec:5:2}, we show that in the case of two subsequent reflections from the structure, the Laplace operator is computed; finally, 
in Subsection~\ref{sec:5:3}, we consider the computing the divergence operator for vector vortex beams possessing a polarization singularity.

\subsection{\label{sec:5:1}Computing a directional derivative}

\begin{figure}
\centering\includegraphics{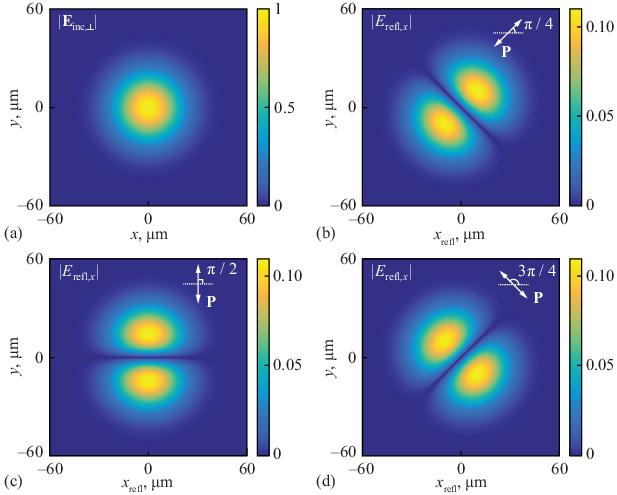}
\caption{\label{fig:4} Incident Gaussian beam $\left| \mat{E}_{{\rm inc}, \bot} \right| = E_0(\mat{r}_{\bot})$~(a) and numerically calculated absolute values of the component $E_{{\rm refl}, x}$ of the reflected beam at different $\varphi$ values: 
$\varphi = \pi/4$~(b), 
$\varphi = \pi/2$~(c), and 
$\varphi = 3\pi/4$~(d).}
\end{figure}

First, let us consider the simplest case of an incident beam linearly polarized along a certain direction $\mat{P}(\varphi ) = \left(\cos\varphi ,\,\sin\varphi \right)$:
\begin{equation}\label{eq:27}
\mat{E}_{{\rm inc}, \bot}(\mat{r}_{\bot},0) 
= \left( 
E_{{\rm inc}, x}(\mat{r}_{\bot}, 0),\;
E_{{\rm inc}, y}(\mat{r}_{\bot}, 0) \right) 
= E_0(\mat{r}_{\bot}) (\cos\varphi,\; \sin\varphi ),
\end{equation}
where $E_0(\mat{r}_{\bot})$ is a given function describing the magnitude of the vector $\mat{E}_{{\rm inc}, \bot}$. 
For such a linearly polarized incident field, the metal-dielectric multilayer of Eq.~\eqref{eq:26} enables calculating the directional derivative of the function $E_0$ (up to a factor $-\ii c_x$):
\begin{equation}\label{eq:28}
E_{{\rm refl},x} \sim \operatorname{div} \mat{E}_{{\rm inc}, \bot} 
= \frac{\partial E_0}{\partial x} \cos\varphi  
+ \frac{\partial E_0}{\partial y} \sin\varphi.
\end{equation}
As an example, Fig.~\ref{fig:4} shows the absolute values of the $x_{\rm refl}$-component of the field $E_{{\rm refl},x}$ reflected from the designed multilayer of Eq.~\eqref{eq:26} for a Gaussian incident beam $E_0(\mat{r}_{\bot}) = \exp \left\{ -(x^2 + y^2)/{\sigma^2} \right\}$ with $\sigma = 20\um$ [Fig.~\ref{fig:4}(a)] at different $\varphi$ values in Eq.~\eqref{eq:27}: 
$\varphi = \pi / 4$ [Fig.~\ref{fig:4}(b)], $\varphi = \pi/2$ [Fig.~\ref{fig:4}(c)], and $\varphi  = 3\pi / 4$ [Fig.~\ref{fig:4}(d)]. These plots were obtained using the ``rigorous'' Eqs.~\eqref{eq:10} and~\eqref{eq:11} and with a high accuracy correspond to the derivatives of the function $E_0(\mat{r}_{\bot})$ along the directions $\mat{P}(\varphi) = ( \cos\varphi ,\, \sin\varphi )$ at the given $\varphi$ values. The normalized root-mean-square (RMS) deviations of the fields shown in Figs.~\ref{fig:4}(b)--\ref{fig:4}(d) from the analytically calculated directional derivatives (after a proper normalization) do not exceed~0.4\%.

Let us note that the linearly polarized beam of Eq.~\eqref{eq:27} with different $\varphi$ values can be generated upon transmission of a circularly polarized beam through a linear polarizer. The required $\varphi$ value and, consequently, the required differentiation direction can be provided by rotating the polarizer axis by the chosen angle $\varphi $ with respect to the $x$ axis [for the presented examples, the orientation directions of the polarizer axis are schematically shown in the upper right parts of Figs.~\ref{fig:4}(b)--\ref{fig:4}(d)]. We believe that the utilization of a multilayer implementing the optical computation of the divergence operator together with a linear polarizer, which can be rotated in order to obtain an incident beam described by Eq.~\eqref{eq:27}, can be promising from the practical point of view for optical image processing and, in particular, for selectively enhancing contours having different directions.

\subsection{\label{sec:5:2}Computing the Laplace operator}

\begin{figure}[bth]
\centering\includegraphics{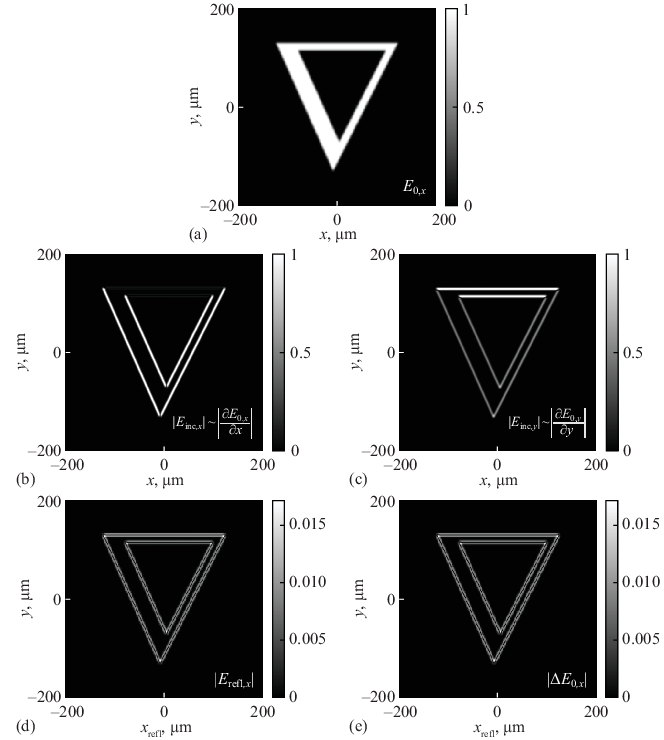}
\caption{\label{fig:5}(a)~Incident field corresponding to the image of the nabla symbol. 
(b),~(c)~Normalized absolute values of the transverse electric field components generated by the designed multilayer. 
(d)~Absolute value of the $E_{{\rm refl}, x}$ component generated upon reflection of the field shown in~(b) and~(c) from the same multilayer. 
(e)~Absolute value of the analytically calculated Laplacian of the input field shown in~(a).}
\end{figure}

Since the divergence of the gradient of a scalar function equals the Laplacian of this function, the designed metal-dielectric multilayer of Eq.~\eqref{eq:26}, being used with the input field 
$\mat{E}_{{\rm inc}, \bot} = ( E_{{\rm inc}, x}, E_{{\rm inc},y} ) 
\sim ( \partial E_{0,x} / \partial x,
\partial E_{0,x} / \partial y)$ corresponding to the gradient of a certain function $E_{0,x}$, enables calculating the Laplacian of this function (up to a constant factor):
\begin{equation}\label{eq:29}
E_{{\rm refl},x} \sim \operatorname{div} \operatorname{grad} E_{0,x} 
= \Delta \,E_{0,x} 
= \frac{\partial^2 E_{0,x}}
			 {\partial x^2} 
+ \frac{\partial^2 E_{0,x}}
			 {\partial y^2}.
\end{equation}
It is important to note that the mentioned input field 
$(E_{{\rm inc}, x}, E_{{\rm inc},y}) \sim (\partial E_{0,x} / \partial x, \partial E_{0,x} / \partial y)$ can also be generated by the designed multilayer of Eq.~\eqref{eq:26} for an incident beam linearly polarized along the $x$ axis [see Eq.~\eqref{eq:21}]. 
As an example, Fig.~\ref{fig:5} shows the $x$ electric field component of the linearly polarized incident beam corresponding to the nabla symbol [Fig.~\ref{fig:5}(a)] and the absolute values of the transverse electric field components formed upon reflection of the incident beam from the designed multilayer of Eq.~\eqref{eq:26} and calculated using the rigorous expressions~\eqref{eq:10} and~\eqref{eq:11} [Figs.~\ref{fig:5}(b) and~\ref{fig:5}(c)].
The obtained field components in Figs.~\ref{fig:5}(b) and~\ref{fig:5}(c) are very close to the exact partial derivatives of the input signal with respect to the spatial variables $x$ and $y$ (the normalized RMS deviations of the fields shown in Figs.~\ref{fig:5}(b) and~\ref{fig:5}(c) from the analytically calculated derivatives do not exceed~1.6\%).

Let then the beam with the field components shown in Figs.~\ref{fig:5}(b) and~\ref{fig:5}(c) generated by the designed multilayer of Eq.~\eqref{eq:26} again impinge on the same structure. In this case, according to Eq.~\eqref{eq:29}, the $x_{\rm refl}$-component of the reflected field will correspond to the Laplacian of the initial input field. In order to demonstrate this fact, Figs.~\ref{fig:5}(d) and~\ref{fig:5}(e) show the absolute value of the $x_{\rm refl}$-component of the reflected field calculated using the rigorous expressions~\eqref{eq:10} and~\eqref{eq:11} [Fig.~\ref{fig:5}(d)] and the normalized absolute value of the analytically calculated Laplace operator of the function $E_{0,x}$ shown in Fig.~\ref{fig:5}(a). The normalized RMS deviation of the absolute value of the~$x_{\rm refl}$-component of the reflected field from the analytically calculated Laplacian amounts to only~1.08\%. Thus, the presented example demonstrates the possibility to compute the Laplace operator using two consecutive reflections of the incident linearly polarized beam from the designed metal-dielectric multilayer.

\subsection{\label{sec:5:3}Computing the divergence operator for vector vortex beams}

In~\cite{27}, the authors considered the optical computation of the divergence operator as a tool for characterizing polarization singularities of vector vortex beams. At the same time, in that work, the authors considered only a particular case of such beams, namely, cylindrical vector Laguerre--Gaussian vortex beams. In this regard, here we present an expression for the divergence of a general-form vector vortex beams (VVB), and then consider a particular example demonstrating the optical computation of the divergence of a Gauss--Bessel VVB using the designed metal-dielectric multilayer of Eq.~\eqref{eq:26}.

For vector vortex beams, the electric field components can be represented in cylindrical coordinates $\rho = (x^2 + y^2)^{1/2}$, $\varphi = \arctan (y/x)$, and $z$ as~\cite{34}
\begin{equation}\label{eq:30}
\begin{aligned}
  E_{{\rm inc},x}(\rho, \varphi, z) &= U(\rho)\xi (z)\cos (n\varphi + \varphi_0),
	\\
  E_{{\rm inc},y}(\rho, \varphi, z) &= U(\rho)\xi (z)\sin (n\varphi + \varphi_0),
\end{aligned}
\end{equation}
where $U(\rho)$ is a certain radial function, 
$\xi(z)$ is an exponential function containing the Gouy phase and describing the evolution of the phase of the beam upon propagation along the $z$ axis, 
and $n$ is the order of the polarization singularity occurring at $r = 0$ (the so-called V-point singularity).
By calculating the divergence of the field of Eq.~\eqref{eq:30} at $z = 0$, we obtain
\begin{equation}\label{eq:31}
\operatorname{div}\mat{E}_{\rm inc}(\rho, \varphi, 0) 
= \xi(0) 
\frac{1}{\rho}\left[ n\, U(\rho) + \rho\frac{{\rm d}U(\rho)}{{\rm d}\rho} \right]
\cos\left( (n - 1)\varphi + \varphi_0 \right).
\end{equation}
From Eq.~\eqref{eq:31}, it follows that the divergence vanishes at the following $2n - 2$ values of the polar angle:
\begin{equation}\label{eq:32}
\tilde \varphi_m = \frac{\pi/2 + m\pi - \varphi_0}{n-1}, \,\, m = 0,...,2n - 3.
\end{equation}

\begin{figure}
\centering\includegraphics{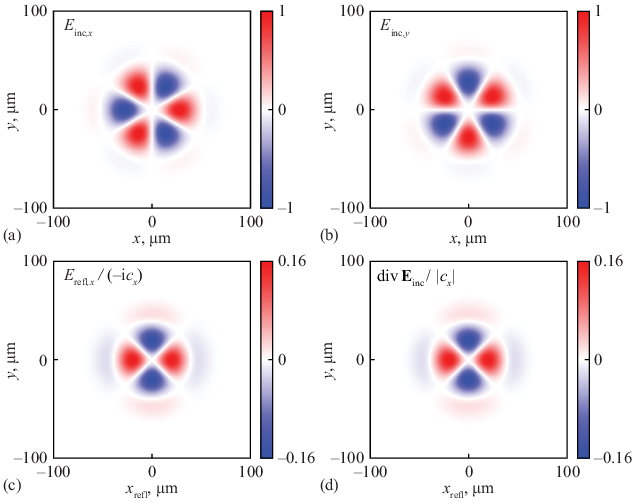}
\caption{\label{fig:6}(a),~(b)~$x$- and $y$- electric field components of the incident beam defined by Eq.~\eqref{eq:33}, (c)~the calculated $x_{\rm refl}$ electric field component of the reflected beam $E_{{\rm refl}, x}$ formed upon reflection of the incident field shown in~(a) and~(b) from the designed metal-dielectric multilayer, and (d)~the analytically calculated divergence of the vector field consisting of the transverse field components of the incident beam.}
\end{figure}

For a numerical demonstration of the optical computation of the divergence operator using the designed metal-dielectric multilayer of Eq.~\eqref{eq:26}, let us consider an incident cylindrical Gauss--Bessel beam having the following form:
\begin{equation}\label{eq:33}
\begin{aligned}
  E_{{\rm inc},x}(\rho,\varphi ,0) &= {\rm J}_s(k_\rho \rho)\exp \left\{ -\rho^2/\sigma^2 \right\} \cos(n\varphi + \varphi_0 ),
	\\
  E_{{\rm inc},y}(\rho,\varphi ,0) &= {\rm J}_s(k_\rho \rho)\exp \left\{ -\rho^2/\sigma^2 \right\} \sin(n\varphi + \varphi_0 ),
\end{aligned}
\end{equation}	
where ${\rm J}_s$ is the Bessel function of the first kind of the $s$-th order, $k_\rho$ is the radial component of the wave vector of the conical wave constituting the Bessel beam, and $\sigma$ is the beam waist of the Gaussian beam.
Figures~\ref{fig:6}(a) and~\ref{fig:6}(b) show in Cartesian coordinates the electric field components $E_{{\rm inc}, x}$, $E_{{\rm inc},y}$ of the beam of Eq.~\eqref{eq:33} at $s = 3$, $n = 3$, $\varphi_0 = 0$, $k_\rho = 0.1\uminv$, and $\sigma  = 30\um$. Figure~\ref{fig:6}(c) shows the field component $E_{{\rm refl},x}$ formed upon reflection of this beam from the designed metal-dielectric multilayer of Eq.~\eqref{eq:26} and calculated using the rigorous expressions~\eqref{eq:10} and~\eqref{eq:11}. Since, according to the ``model'' Eq.~\eqref{eq:19}, the computation of the divergence is performed up to the factor $-\ii c_x$, the field component $E_{{\rm refl},x}$ shown in Fig.~\ref{fig:6}(c) is normalized by this factor. For the sake of comparison, Fig.~\ref{fig:6}(d) shows the analytically calculated divergence of the field of Eq.~\eqref{eq:33}. It is evident that the reflected field [Fig.~\ref{fig:6}(c)] with a high accuracy coincides with the analytically calculated divergence [Fig.~\ref{fig:6}(d)]; the normalized RMS deviation of the calculated field $E_{{\rm refl},x} / ( -\ii c_x)$ from the analytically calculated divergence does not exceed~1.3\%. One can also see that the component $E_{{\rm refl},x}$ vanishes at $\varphi = {\tilde \varphi}_m = \pi / 4 + m \pi  / 2,\,\,m = 0,...,3$, which, according to Eq.~\eqref{eq:32}, corresponds to a polarization singularity of the third order ($n = 3$).

\section{\label{sec:6}Conclusion}

In the present work, we demonstrated the possibility of the optical computation of the divergence operator using a layered structure. For this, we derived a vectorial transfer function describing the transformation of the electric field components of an incident three-dimensional light beam that occurs upon reflection from the structure.
By analyzing this TF, we obtained the conditions required for computing the divergence of a two-dimensional vector field, which is composed of the transverse electric field components of the incident beam. 
The divergence is computed in one of the transverse electric field components of the reflected beam.
We demonstrated that in the particular case of a linearly polarized incident beam, the layered structure optically implementing the divergence operator also makes it possible to compute the gradient of the nonzero transverse electric field component of the incident beam in the transverse components of the reflected beam. As an example of a layered structure computing the divergence operator, we proposed a four-layer metal-dielectric structure operating in the oblique incidence geometry.
The presented numerical simulation results of the designed metal-dielectric multilayer demonstrate the optical computation of the divergence operator with high quality and show the possibility of using this structure for the optical computation of directional derivatives, and for computing the gradient and the Laplace operator. We believe that the obtained results may find application in novel analog optical computing and optical information processing systems.

\begin{backmatter}

\bmsection{Acknowledgments}
This work was funded by the Russian Science Foundation (project 24-12-00028; theoretical description of the computation of the divergence operator, design and numerical investigation of the metal-dielectric multilayer optically computing the divergence) and performed within the state assignment to the NRC ``Kurchatov Institute'' (implementation of the software for simulating the diffraction of optical beams on layered structures).



\end{backmatter}

\bibliography{div_mdm}

\end{document}